\documentclass[
reprint, 
amsmath,amssymb,
aps,pra,
]{revtex4-2}

\usepackage{graphicx}  
\usepackage{dcolumn}   
\usepackage{bm}        
\usepackage{amsthm}
\usepackage{braket}
\usepackage{balance}

\begin{document}
	
	\preprint{}
	
	\title{\textbf{Noise Resilience of Spin Quantum Battery in the presence of DM Interactions} 
	}%

	\author{B. Vigneshwar,}
	\author{R. Sankaranarayanan }%
	\affiliation{ Department of Physics, National Institute of Technology, Tiruchirappalli 620015, TamilNadu, India.\\}

	
	\begin{abstract}
		Quantum batteries utilize quantum effects to enhance energy storage and work extraction, offering promising avenues for nanoscale energy applications. However, environmental noise poses a significant challenge by degrading stored energy. For a single qubit, we show that amplitude damping and bit-flip noises lead to ergotropy loss, while phase-flip noise permits partial preservation of work. Extending to a two-qubit Heisenberg XYZ model with Dzyaloshinsky-Moria interaction (DMI), we identify a critical interaction strength that enhances energy retention. We show that strong DMI and initial coherence protects ergotropy even under repeated noise applications, highlighting chiral spin interactions as a resource for noise-resilient quantum batteries.
	\end{abstract}

	\maketitle

	\section{Introduction}
	In recent years, there has been growing interest in leveraging quantum phenomena to advance next-generation technologies~\cite{acin2018quantum,vinjanampathy2016quantum,campaioli2024colloquium}. This pursuit is often framed as achieving quantum supremacy in energy manipulation and storage, fueled by parallel developments in quantum thermodynamics and quantum information science~\cite{kosloff2013quantum,pekola2015towards,galindo2002information}. In this context, quantum batteries have emerged as promising platforms for exploiting quantum effects to enhance performance~\cite{alicki2013entanglement,rossini2020quantum,andolina2019extractable,julia2020bounds,mazzoncini2023optimal}.
	The concept of ergotropy, introduced in~\cite{allahverdyan2004maximal}, quantifies the maximum extractable work from a quantum system. Quantum advantages in the charging process have been demonstrated by studying ergotropy through mechanisms such as quantum speed limits, entangling unitary operations, non-reciprocity, topological and many-body interactions~\cite{campaioli2017enhancing,francica2020quantum,rossini2019many,ahmadi2024nonreciprocal,song2024remote,lu2025topological}. However, recent experimental implementations have highlighted the critical importance of stability under environmental noise to ensure reliable energy recovery~\cite{quach2022superabsorption,hu2022optimal,maillette2023experimental,razzoli2025cyclic,gemme2022ibm}.
	
	In particular, it has been shown that in a charger-mediated setup interacting with a Markovian environment, the ergotropy of a qubit battery vanishes at steady state when only thermal resources are available~\cite{andolina2018charger}. Consequently, in the presence of amplitude damping noise, quantum coherence has been found to enhance energy retention, and the use of nonlocal correlations further improves the efficiency of work extraction~\cite{tirone2025quantum,tirone2023work}. Furthermore, dephasing-assisted protocols have shown that noise, rather than being purely detrimental, can enhance charging dynamics~\cite{shastri2025dephasing}, underscoring the need to understand the nuanced effects of noise on the energy storage.
	The Dzyaloshinsky–Moriya interaction (DMI) is an anti-symmetric exchange interaction that emerges due to spin–orbit coupling and broken inversion symmetry in certain crystal structures \cite{choi2019colloquium}. Originally introduced to explain weak ferromagnetism in antiferromagnetic systems~\cite{moriya1960new,dzyaloshinsky1958thermodynamic}, DMI favors chiral spin arrangements that are not captured by conventional isotropic exchange models. The presence of DMI can significantly modify energy level structures and spin dynamics \cite{rahman2024effect}, making it a valuable tool in engineered quantum systems for manipulating coherence and population asymmetry.
	
	While prior studies have mainly focused on the effects of environmental noise during the charging and work extraction phases~\cite{zhao2021quantum,barra2019dissipative}, the resilience of stored ergotropy under repeated noise exposure during the post-charging storage phase warrants further investigations. In this work, we address this gap by extending the protocol proposed in~\cite{tirone2025quantum} to systematically investigate how different noise channels, upon repeated application, impact the ergotropy of a charged quantum battery during storage.

	Our work begins by analyzing a single-qubit battery under three canonical types of noise: phase-flip, bit-flip, and amplitude damping. We find that under repeated applications of phase-flip channels, fully charged initial states maintain nonzero ergotropy in the asymptotic limit. This indicates that dephasing, while beneficial for charging speed, severely restricts partial charge retention during storage. In contrast, bit-flip and amplitude damping noises lead to rapid, irreversible ergotropy decay. Having established baseline behavior in the single-qubit case, we then explore a two-qubit battery modeled by the anisotropic Heisenberg XYZ spin system with an additional DMI term along the $z$-direction. In the noiseless scenario, we identify three regimes: $R_1$, $R_2$, and $R_3$ based on critical values of DMI strength. These regimes show distinct behaviors in energy gap structure and ergotropy, with $R_2$ and $R_3$ demonstrating significantly enhanced work storage capacity due to the asymmetric interactions introduced by DMI.
	Subsequently we extend the analysis on the effect of noise during the storage phase to the two-qubit system. In the repeated application of amplitude damping noise, we observe that ergotropy decays toward zero in $R_1$ but stabilizes at nonzero values in $R_2$ and $R_3$, owing to population dynamics of amplitude damping favoring the asymmetry in the DMI-dominating regimes. Under the repeated application of the bit-flip noise, the stability of ergotropy demonstrates critical dependence on the initial state's coherence, correlating residual anti-diagonal elements with asymptotic ergotropy. This coherence-dependence leads to intriguing reversals in the critical regimes where $R_2$ may outperform $R_3$ when initial coherence is higher, despite $R_3$ being optimal in the noiseless case.  Hence by identifying the regimes where ergotropy persists and clarifying the roles of coherence, our results contribute to robust design architectures for building noise-resilient quantum energy storage devices.
	
	Our work is structured as follows. We begin by defining the single-qubit model and noise protocols in Sec. \ref{sec2}. The two-qubit model is introduced and the critical regimes are studied in Sec. \ref{sec3}. Section \ref{sec4} delves into noise protocols in the two-qubit model. Finally, the key results are summarized in Sec \ref{sec5}.


	\section{Single-Qubit Model} \label{sec2}
	
	We begin by considering the Hamiltonian of a two-level quantum battery, expressed as
	\begin{equation}
		\hat{H} = \frac{h_0}{2}(\hat{\sigma}_z + \hat{\mathbb{I}}) = h_0 \hat{\sigma}_+ \hat{\sigma}_-,
		\label{1e}
	\end{equation}
	where $\hat{\sigma}_{\pm} = \frac{1}{2}(\hat{\sigma}_x \pm i \hat{\sigma}_y)$ are the ladder operators constructed from the Pauli matrices, and $h_0$ denotes the energy level spacing, which is set to unity for simplicity. The system is initially prepared in the ground state of the Hamiltonian, $\hat{\rho}_0 = |1\rangle \langle 1|$.
	
	The battery is charged through a direct charging protocol \cite{campaioli2024colloquium,le2018spin,ghosh2020enhancement}, wherein an external driving field described by $\hat{H}_f = \omega \hat{\sigma}_x$ is applied. The field strength $\omega$ is assumed to be much larger than the energy scale of $\hat{H}$, allowing the effective charging Hamiltonian to be approximated as
	\begin{equation}
		\hat{H}_c = \kappa(\tau) \hat{H}_f + \hat{H} \approx \kappa(\tau) \omega \hat{\sigma}_x.
		\label{1ae}
	\end{equation}
	 The time-dependent switching function $\kappa(\tau)$ controls the activation of the charging field, with $\kappa(\tau) = 1$ during the charging interval $\tau \in [0, t]$ and $\kappa(\tau) = 0$ otherwise.
	
	Following this protocol, the state of the system evolves unitarily under the action of $\hat{H}_c$, leading to the time-evolved state $\hat{\rho}(t) = \hat{U}_c \hat{\rho}_0 \hat{U}_c^\dagger$, where $\hat{U}_c = \exp(-i \hat{H}_c t)$ with $\hbar = 1$. The corresponding density matrix of the charged state takes the form
	\begin{equation}
		\hat{\rho}(t) =
		\begin{pmatrix}
			\sin^2(\omega t) & -\frac{i}{2} \sin(2 \omega t) \\
			\frac{i}{2} \sin(2 \omega t) & \cos^2(\omega t)
		\end{pmatrix}.
		\label{2e}
	\end{equation}
	\subsection{Ergotropy}
	
	Ergotropy quantifies the maximum extractable work from a quantum system by cyclic unitary operations \cite{allahverdyan2004maximal}. For a given quantum state $\hat{\rho}$ with spectral decomposition $\hat{\rho} = \sum_n r_n |r_n\rangle \langle r_n|$, and a Hamiltonian $\hat{H} = \sum_n e_n |e_n\rangle \langle e_n|$, where the eigenvalues are ordered such that $r_1 \ge r_2 \ge \cdots$ and $e_1 \le e_2 \le \cdots$, the passive state $\hat{\rho}^p$ associated with $\hat{H}$ is defined as $\hat{\rho}^p = \sum_n r_n |e_n\rangle \langle e_n|$. The ergotropy is then given by
	\begin{equation}
		\xi = \text{Tr}[\hat{H} (\hat{\rho} - \hat{\rho}^p)] = \sum_{m,n} r_m e_n \left( |\langle r_m | e_n \rangle|^2 - \delta_{mn} \right).
		\label{3e}
	\end{equation}
	For the single-qubit system described above, the ergotropy can alternatively be expressed in terms of expectation values of Pauli operators \cite{andolina2018charger} as
	\begin{equation}
		\xi = \frac{1}{2} \left( \langle \hat{\sigma}_z \rangle + \sqrt{ \langle \hat{\sigma}_z \rangle^2 + 4 \langle \hat{\sigma}_+ \rangle \langle \hat{\sigma}_- \rangle } \right),
		\label{4e}
	\end{equation}
	where $\langle \hat{o} \rangle = \text{Tr}[\hat{\rho} \hat{o}]$ denotes the expectation value of operator $\hat{o}$.
	
	\subsection{Noisy Channels}
	\begin{figure*}[ht]
		\centering
		\includegraphics[width=1\linewidth]{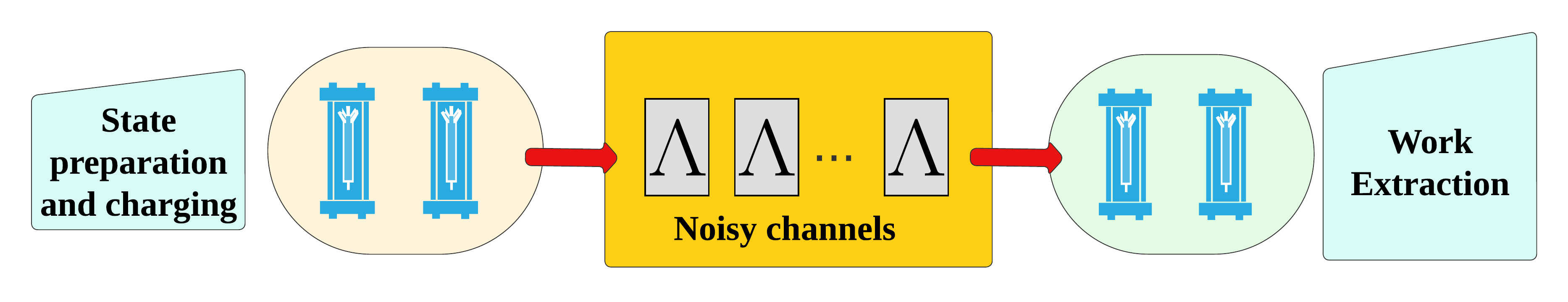}
		\caption{Schematic diagram illustrating the interaction of the charged state with sequential noisy channels.}
		\label{p1}
	\end{figure*}
	In realistic scenarios, quantum systems are inevitably affected by noise. Without fault-tolerant architectures, environmental interactions must be incorporated to accurately model quantum dynamics. General quantum operations, including noise effects, can be described using the Kraus operator formalism \cite{nielsen2010quantum}. In this representation, a completely positive trace-preserving (CPTP) map $\Lambda$ acting on a state $\hat{\rho}$ takes the form
	\begin{equation}
		\Lambda(\hat{\rho}) = \sum_i \hat{K}_i \hat{\rho} \hat{K}_i^\dagger,
		\label{5e}
	\end{equation}
	where $\hat{K}_i$ are the Kraus operators satisfying the completeness relation $\sum_i \hat{K}_i^\dagger \hat{K}_i = \mathbb{I}$. Let $\Lambda^2(\hat{\rho}) = \Lambda(\Lambda(\hat{\rho}))$ denote the state after two sequential applications of the noise channel. More generally, we define $\Lambda^N(\hat{\rho})$ as the quantum state resulting from $N$ sequential applications of the same CPTP map $\Lambda$.
	In this study, we investigate the effect of noisy channels on the ergotropy of the charged quantum battery during the storage phase i.e., the period after charging and prior to work extraction. While prior studies have analyzed environmental effects during the charging process via master equation approaches \cite{farina2019charger}, it is equally important to understand how noise influences the stored energy during transport or delay before extraction. To this end, we propose an extension of the protocol in \cite{tirone2025quantum} wherein the charged state is subjected to a repeated sequence of noisy quantum channels. The schematic of the protocol is shown in Fig.~\ref{p1}, and the procedure is as follows:
	
	\begin{itemize}
		\item The system is initialized in the ground state and charged using the external field.
		\item The charged state is then sequentially passed through a series of noisy channels $\Lambda$.
		\item Work is extracted via a unitary transformation that maps the final state to its passive counterpart.
	\end{itemize}
	A desirable feature of a quantum battery is the robustness of ergotropy against successive noisy operations, as this directly impacts the reliability of stored energy.

	\subsubsection{Phase-Flip (PF)}
	
	Phase-flip noise is a type of decoherence that selectively alters the relative phase between the computational basis states. Specifically, it flips the phase of the $\ket{1}$ state while leaving the $\ket{0}$ state unchanged. The corresponding Kraus operators are given by
	\begin{equation}
		\hat{K}_1 = \sqrt{1 - p} \, \hat{\sigma}_z, \quad \hat{K}_2 = \sqrt{p} \, \mathbb{I},
	\end{equation}
	where $p$ denotes the probability that no phase flip occurs, while $1 - p$ represents the probability of a phase flip.
	Let $N$ denote the number of sequential applications of the phase-flip channel. The evolution of the system under repeated phase-flip noise can be analyzed by applying the channel $N$ times to the charged state. The resulting ergotropy, computed using Eq.~\eqref{4e}, is given by
	\begin{equation}
		\xi = \frac{1}{2} \left( -\cos(2 \omega t) + \sqrt{ \cos^2(2\omega t) + (2p - 1)^{2N} \sin^2(2\omega t) } \right).
		\label{7e}
	\end{equation}
	Additional analytical details in deriving Eq.~\eqref{7e} are provided in Appendix~\ref{appa}. In the asymptotic limit $N \to \infty$, the expression simplifies to
	\begin{equation}
		\xi =
		\begin{cases}
			0 & \text{if } \cos(2\omega t) > 0, \\
			-\cos(2\omega t) & \text{otherwise},
		\end{cases}
		\label{8e}
	\end{equation}
	indicating that, regardless of the probability $p$, the ergotropy retains a time-dependent structure even in the infinite-channel limit.
	This result highlights that the asymptotic behavior of ergotropy under phase-flip noise becomes independent of the specific noise parameter $p$, with the function narrowing symmetrically around the maximum ergotropy values as illustrated in Fig. \ref{p2}(a). 
	
	\subsubsection{Bit-Flip (BF)}
	
\begin{figure*}[ht]
		\centering
		\includegraphics[width=1\linewidth]{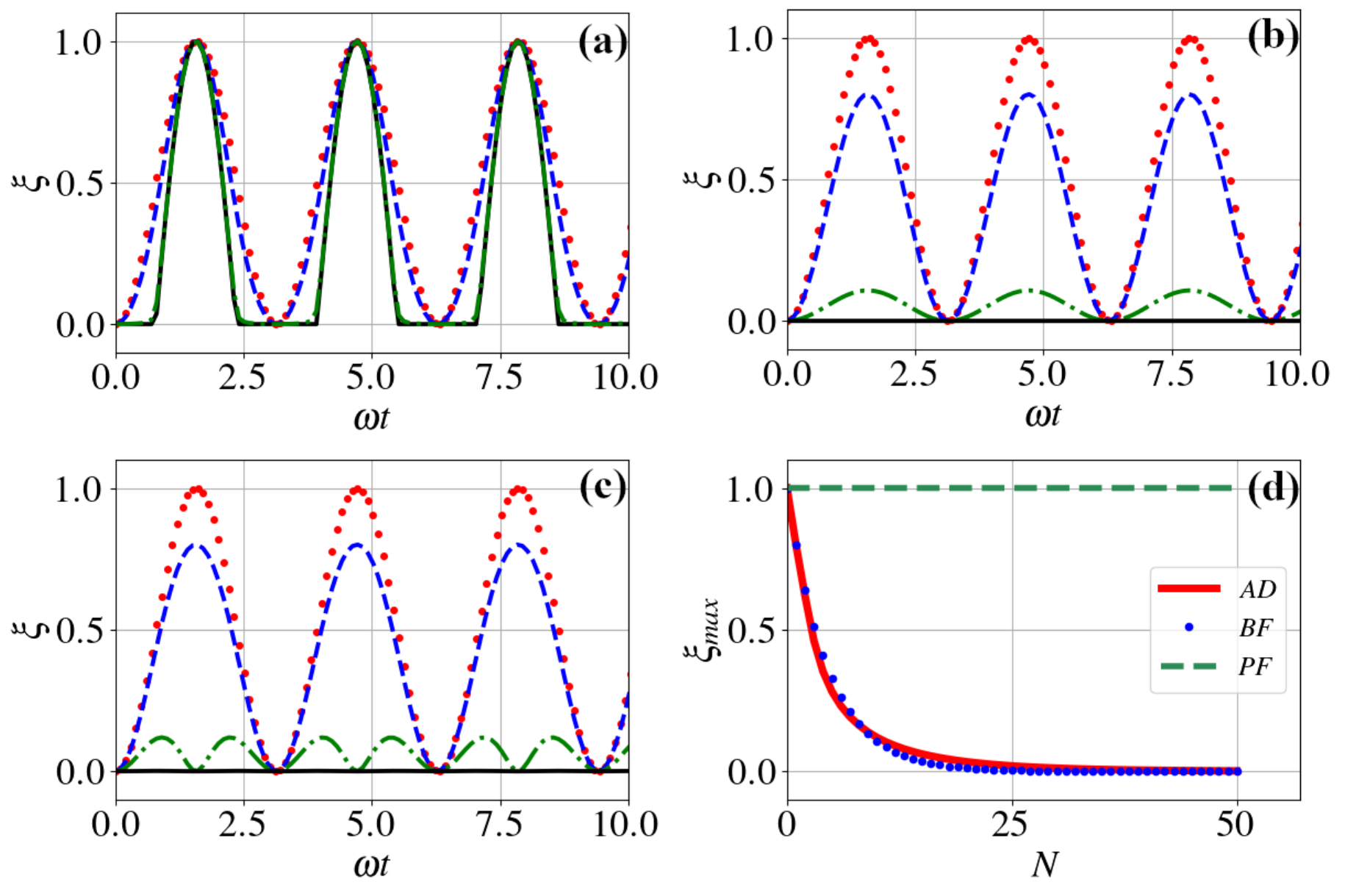}
		\caption{Ergotropy $\xi$ is after application of $N=0$ (red dots), $N=1$ (blue dashed lines), $N=10$ (green dash-dotted lines), and $N=50$ (black solid lines), respectively for the case of (a) phase-flip noise and (b) bit-flip noise, and (c) amplitude damping noise with $p=0.1$. (d) The maximum of ergotropy $\xi_{max}$ for increasing number of channels $N$ for the three noise scenarios.}
		\label{p2}
	\end{figure*}

	Bit-flip noise describes a channel in which the system undergoes a logical bit flip with a certain probability. The corresponding Kraus operators are given by
	\begin{equation}
		\hat{K}_1 = \sqrt{1 - p} \, \hat{\sigma}_x, \quad \hat{K}_2 = \sqrt{p} \, \mathbb{I},
	\end{equation}
	where $p$ denotes the probability that the state remains unchanged, and $1 - p$ represents the probability of a bit-flipping.
	After $N$ sequential applications of the noise channel, the ergotropy of the system can be evaluated using Eq.~\eqref{4e} (refer Appendix \ref{appa}). The result takes the form
	\begin{equation}
		\xi =
		\begin{cases}
			-\dfrac{1}{2} (2p - 1)^N (\cos(2 \omega t) - 1), & \text{if } p > 0.5, \\
			0, & \text{if } p = 0.5, \\
			-\dfrac{1}{2} (2p - 1)^N \left( \cos(2 \omega t) + (-1)^{N + 1} \right), & \text{if } p < 0.5.
		\end{cases}
		\label{6e}
	\end{equation}
	Noting that $0 <(2p - 1)<1$ for all $p \in (0,1)$, it follows that in the asymptotic limit $N \to \infty$, $(2p - 1)^N \to 0$ and hence the ergotropy also decays to zero.
	\begin{equation}
		\lim_{N \to \infty} \xi = 0.
	\end{equation}
	This behavior indicates that, unlike the phase-flip channel where residual ergotropy may persist under certain conditions, bit-flip noise uniformly degrades the stored energy over repeated applications. Consequently, the bit-flip channel acts as a more severe form of decoherence from the perspective of energy retention, rapidly eroding the system’s ability to store extractable work. This result is demonstrated in Fig. \ref{p2}(b), where the oscillating ergotropy decays with increasing $N$.
	\subsubsection{Amplitude Damping (AD)}
	
	Amplitude damping noise models energy dissipation from a quantum system to its environment, such as spontaneous emission in a two-level atom. It leads to irreversible loss of excitations, thereby reducing the system’s capacity to store useful energy. The corresponding Kraus operators for a single-qubit amplitude damping channel are given by
	\begin{equation}
		\hat{K}_1 = 
		\begin{pmatrix}
			\sqrt{1 - p} & 0 \\
			0 & 1
		\end{pmatrix}, \quad
		\hat{K}_2 = 
		\begin{pmatrix}
			0 & 0 \\
			\sqrt{p} & 0
		\end{pmatrix},
	\end{equation}
	where $p$ is the damping parameter representing the probability of decay from the excited state to the ground state.
	After $N$ sequential applications of the noise channel, the ergotropy of the system is given by (see Appendix \ref{appa})
	\begin{equation}
		\xi = \frac{1}{2} \left( \Gamma_N + \sqrt{ \Gamma_N^2 + (1 - p)^N \sin^2(2 \omega t) } \right),
		\label{9e}
	\end{equation}
	where
	\begin{align}
		\Gamma_N ={}& (1 - p)^N \sin^2(\omega t) \nonumber \\
		& - \left( \cos^2(\omega t) + \sum_{j = 0}^{N - 1} p (1 - p)^j \sin^2(\omega t) \right).
	\end{align}
	
	
	In the limit $N \to \infty$, it can be shown that $\Gamma_N \to -1$ and $\xi \to 0$, indicating complete degradation of extractable work under repeated amplitude damping interactions. Figure~\ref{p2}(c) shows that, as in the bit-flip case, the ergotropy decays monotonically with $N$. This reflects the inherently dissipative nature of amplitude damping, which irreversibly transfers energy from the system to the environment, leaving the battery in its ground state in the asymptotic limit. Additionally, Fig.~\ref{p2}(d) depicts the maximum ergotropy $\xi_{\text{max}}$ as a function of the number of noise channels. While $\xi_{\text{max}}$ asymptotically decays to zero for both bit-flip and amplitude damping channels, phase-flip noise uniquely preserves a finite value as expected.
	While dephasing-assisted protocols have been shown to facilitate fast charging by optimizing the interplay between coherence and dissipation \cite{shastri2025dephasing}, our results emphasize a complementary aspect: the long-term stability of stored energy under dephasing noise. In particular, we find that in the asymptotic limit of repeated phase-flip interactions, ergotropy is preserved for fully charged states. This implies that, although dephasing can accelerate energy transfer during charging \cite{shastri2025dephasing}, it imposes strict limitations on the persistence of partial charge during storage. Therefore, an optimal battery protocol must balance the advantages of dephasing-enhanced charging with the ergotropy degradation associated with prolonged exposure to phase noise.
	In contrast, amplitude damping noise and bit-flip noise lead to a rapid and irreversible decay of ergotropy. By focusing on how these noise processes impact the post-charging stability of ergotropy across different noise environments, our results provides a unique perspective on quantum battery performance under realistic storage and transport conditions.
	
	\section{Two-Qubit Heisenberg System} \label{sec3}
	
	Having examined the impact of sequential noise on the ergotropy of a single qubit, we now extend our investigation to a two-qubit interacting system. In particular, we consider the anisotropic Heisenberg XYZ model augmented with Dzyaloshinsky-Moriya interaction (DMI).
	
	\subsection{Model}
	
	The Hamiltonian of the system is given by
	\begin{align}
		\hat{H} =\, & h_0 \left( \hat{\sigma}_+^{(1)} \hat{\sigma}_-^{(1)} + \hat{\sigma}_+^{(2)} \hat{\sigma}_-^{(2)} \right) \nonumber
		\\& +\frac{J}{2} \left[ (1 + \gamma) \hat{\sigma}_x^{(1)} \hat{\sigma}_x^{(2)} + (1 - \gamma) \hat{\sigma}_y^{(1)} \hat{\sigma}_y^{(2)} \right] \nonumber \\
		& + \frac{J_z}{2} \hat{\sigma}_z^{(1)} \hat{\sigma}_z^{(2)} 
		+ \frac{D}{2} \left( \hat{\sigma}_x^{(1)} \hat{\sigma}_y^{(2)} - \hat{\sigma}_y^{(1)} \hat{\sigma}_x^{(2)} \right),
		\label{e1}
	\end{align}
	where $\hat{\sigma}_i^{(k)}$ ($i = x, y, z$) denotes the $i$th Pauli operator acting on the $k$th qubit. The parameter $h_0$ characterizes the local energy level spacing due to the external field, $J$ and $J_z$ quantify the Heisenberg-type spin-spin interactions, $\gamma$ introduces anisotropy in the $xy$-plane, and $D$ represents the strength of the Dzyaloshinsky-Moriya interaction, which induces spin canting through antisymmetric exchange.
	
	The eigenvalues and eigenvectors of the Hamiltonian~\eqref{e1} are given by
	\begin{subequations}
		\begin{align}
			e_{1,2} &= h_0 \pm \sqrt{J^2 + D^2} - \frac{J_z}{2}, \\
			|e_{1,2}\rangle &= \frac{1}{\sqrt{|c_{1,2}|^2 + 1}} \left( 0,\, c_{1,2},\, 1,\, 0 \right)^\intercal, \\
			e_{3,4} &= h_0 \pm \sqrt{h_0^2 + J^2 \gamma^2} + \frac{J_z}{2}, \\
			|e_{3,4}\rangle &= \frac{1}{\sqrt{|c_{3,4}|^2 + 1}} \left( c_{3,4},\, 0,\, 0,\, 1 \right)^\intercal,
		\end{align}
		\label{11e}
	\end{subequations}
	where the coefficients are defined as
	\begin{align}
		c_{1,2} &= \pm \frac{J + i D}{\sqrt{J^2 + D^2}}, \quad
		c_{3,4} = \frac{h \pm \sqrt{h^2 + J^2 \gamma^2}}{J \gamma}.
	\end{align}
	To explore the energy storage characteristics of this system, we fix the field strength to $h_0 = 1$ and systematically study the interaction parameters $J$, $J_z$, $\gamma$, and $D$. This allows us to classify different regimes for DMI assisted ergotropy enhancement.
	
	\subsection{Critical Values and DMI-Dominated Regime}
	\begin{figure*}[ht]
		\centering
		\includegraphics[width=1\linewidth]{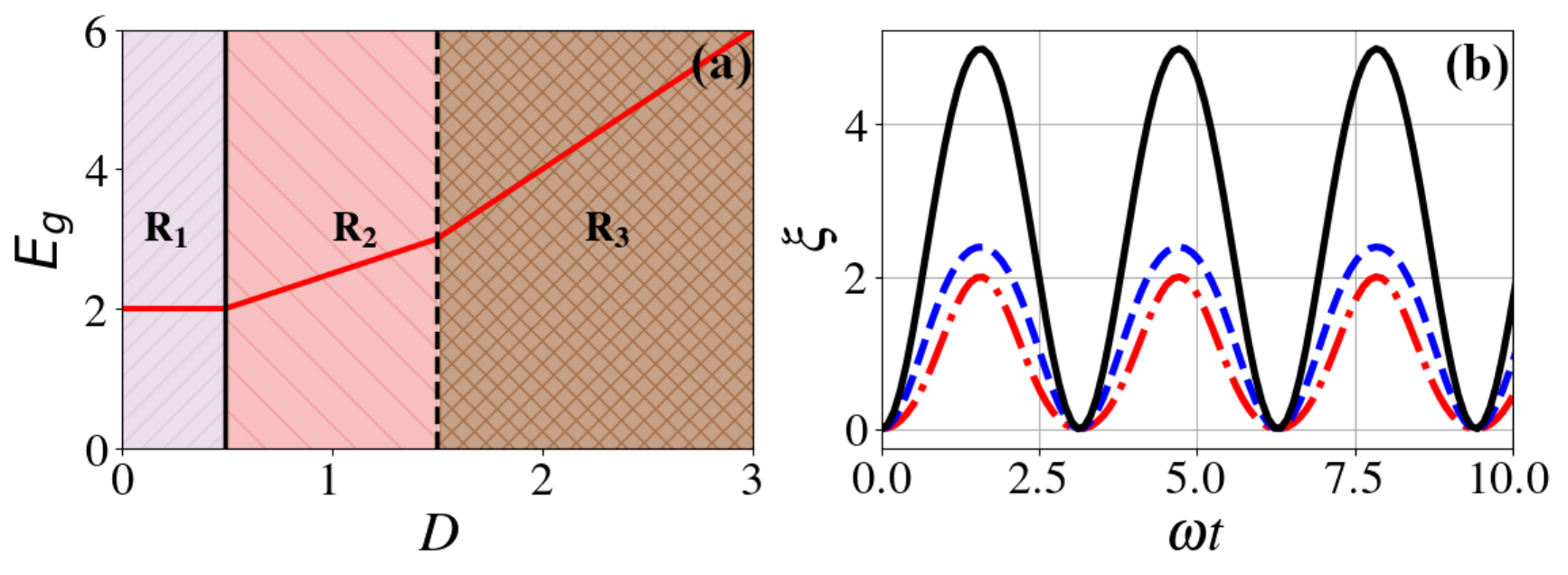}
		\caption{(a) The energy gap $E_g$ as a function of DMI strength $D$. The solid black line denotes $D=D_c$ and the dotted line denotes $D=D_c'$. (b) The ergotropy $\xi$ is shown for the three regions $R_1$, $R_2$, and $R_3$ for the DMI strength $D=0.3$ (red dash-dotted line), $D=1.2$ (blue dashed line), and $D=2.5$ (black solid line), respectively. Other parameter values are $J=0.1$, $J_z=0.5$, and $\gamma=0.2$. }
		\label{p3}
	\end{figure*}
	Throughout this study, we consider a regime characterized by ferromagnetic alignment along the $z$-axis, enforced by the condition $J_z > 0$. In this case, the ground state of the system corresponds to either $e_2$ or $e_4$, depending on the competition between the symmetric and anti-symmetric spin interactions. To delineate the regions of dominance, we analyze the inequality $e_2 < e_4$ which yields the condition
	\begin{equation}
		\sqrt{D^2 + J^2} > \sqrt{1 + J^2 \gamma^2} - J_z.
		\label{12e}
	\end{equation}
	Similarly, to determine the transition in the highest excited state, we examine the inequality $e_1 > e_3$, resulting in the condition
	\begin{equation}
		\sqrt{D^2 + J^2} > \sqrt{1 + J^2 \gamma^2} + J_z.
		\label{12ae}
	\end{equation}
	Based on the inequalities defined above, we identify two critical values for the DMI strength: $D_c$ and $D_c'$, where
	\begin{itemize}
		\item $D > D_c$ denotes the parameter range in which only the ground state transitions occur, satisfying inequality~\eqref{12e}, and
		\item $D > D_c'$ corresponds to the regime where both the ground and excited state transitions occur, satisfying inequalities~\eqref{12e} and \eqref{12ae}.
	\end{itemize}
	This classification naturally leads to three distinct interaction regions:
	\begin{itemize}
		\item Region 1 ($R_1$): $D < D_c$ — Heisenberg-dominated ,
		\item Region 2 ($R_2$): $D_c < D < D_c'$ — partial DMI influence ,
		\item Region 3 ($R_3$): $D > D_c'$ —  DMI-dominated.
	\end{itemize}
	
	To characterize the storage capacity of the system, we define the energy gap
	\begin{equation}
		E_g = \max \{e_i\} - \min \{e_i\},
	\end{equation}
	which sets an upper bound on the ergotropy achievable via unitary work extraction. For the Hamiltonian in Eq.~\eqref{e1}, $E_g$ remains constant with increasing $D$ in Region $R_1$. Upon entering Region $R_2$, the energy gap begins to increase linearly with $D$, reflecting the onset of DMI-dominated behavior. In Region $R_3$, this growth continues with a larger slope, as shown in Fig.~\ref{p3}(a), indicating enhanced capacity for energy storage.
	Importantly, the two critical points $D_c$ and $D_c'$ are distinct only when $J_z \neq 0$, highlighting the essential role of symmetric spin alignment in modulating the energy landscape under the influence of anti-symmetric DMI.

	 \subsection{Ergotropy of the two Qubit battery}
	 
	 To evaluate the energy storage capabilities of the interacting two-qubit system, we consider a charging protocol governed by a strong external field,
	 \begin{equation}
	 	\hat{H}_f = \omega \left( \hat{\sigma}_x^{(1)} + \hat{\sigma}_x^{(2)} \right).
	 \end{equation}
	 The choice of $\hat{H}_f$ mirrors that in Eq.~\eqref{1ae}, with the driving strength $\omega$ assumed to be much larger than the intrinsic interaction parameters in Eq.~\eqref{e1}. Under this assumption, the effective charging Hamiltonian simplifies to
	 \begin{equation}
	 	\hat{H}_c = \kappa(\tau) \hat{H}_f + \hat{H} \approx \kappa(\tau) \omega \left( \hat{\sigma}_x^{(1)} + \hat{\sigma}_x^{(2)} \right),
	 	\label{13ae}
	 \end{equation}
	 where $\kappa(\tau)$ is the time-dependent switching function defined below Eq.~\eqref{1ae}. The system is initialized in its ground state, which depends on the DMI strength.
	 \begin{equation}
	 	\hat{\rho}_0 = 
	 	\begin{cases}
	 		\left| e_4 \right\rangle \left\langle e_4 \right|, & \text{if } D \leq D_c, \\
	 		\left| e_2 \right\rangle \left\langle e_2 \right|, & \text{if } D > D_c.
	 	\end{cases}
	 	\label{13e}
	 \end{equation}
	 The state of the system at time $t$ following unitary evolution is given by
	 \begin{equation}
	 	\hat{\rho}(t) = \hat{U}_c \hat{\rho}_0 \hat{U}_c^\dagger,
	 \end{equation}
	 where the unitary operator under the strong field approximation reads
	 \begin{equation}
	 	\hat{U}_c = e^{-i \hat{H}_c t} = 
	 	\frac{1}{2} \begin{pmatrix}
	 		a + 1 & b & b & a - 1 \\
	 		b & a + 1 & a - 1 & b \\
	 		b & a - 1 & a + 1 & b \\
	 		a - 1 & b & b & a + 1
	 	\end{pmatrix},
	 	\label{14e}
	 \end{equation}
	 with $a = \cos(2\omega t)$ and $b = -i \sin(2\omega t)$.
	 Since the initial state $\hat{\rho}_0$ is passive with respect to the system Hamiltonian $\hat{H}$, the ergotropy is equal to the net stored energy under unitary evolution \cite{konar2022quantum}. It is therefore computed as
	 \begin{equation}
	 	\xi = \text{Tr}[\hat{H} \hat{\rho}(t)] - \text{Tr}[\hat{H} \hat{\rho}_0].
	 \end{equation}
	 \subsubsection{ Region $R_1$ }
	 
	 For $D < D_c$, the ground state is $\left| e_4 \right\rangle$, and the ergotropy becomes
	 \begin{equation}
	 	\xi = \frac{\phi_0 (2h + \frac{J_z}{2}) + 2 \phi_1 J\gamma + 2 \phi_2 (h + J - \frac{J_z}{2}) + \phi_3 \frac{J_z}{2} - \phi_4}{4(c_4^2 + 1)},
	 	\label{15e}
	 \end{equation}
	 where the coefficients are defined as
	 \begin{subequations}
	 	\begin{align}
	 		\phi_0 &= (a + 1)^2 c_4^2 + 2c_4 (a^2 - 1) + (a - 1)^2, \\
	 		\phi_1 &= (a^2 - 1) c_4^2 + 2c_4 (a^2 + 1) + a^2 - 1, \\
	 		\phi_2 &= |b|^2 (c_4 + 1)^2, \\
	 		\phi_3 &= (a - 1)^2 c_4^2 + 2c_4 (a^2 - 1) + (a + 1)^2, \\
	 		\phi_4 &= 4 c_4^2 \left(2h_0 + \frac{J_z}{2}\right) + 8 J\gamma c_4 + 2J_z.
	 	\end{align}
	 \end{subequations}
	 Particularly, $\xi$ is independent of the DMI strength $D$ in this region. Hence, energy storage in $R_1$ is governed entirely by the symmetric spin interactions.
	 
	 \subsubsection{ Regions $R_2$ and $R_3$}
	 
	 For $D > D_c$, the ground state switches to $\left| e_2 \right\rangle$, and the ergotropy takes the form
	 \begin{align}
	 	\xi = \frac{1}{4(1 + |c_2|^2)} \Big[ & \eta_0 (2h_0 + J_z + 2J\gamma) + (\eta_1 + \eta_2)\left(h_0 - \frac{J_z}{2}\right) \nonumber \\
	 	& + 2 \operatorname{Re}(\eta_3 (J - iD)) - \eta_4 \Big],
	 	\label{16e}
	 \end{align}
	 with coefficients defined as
	 \begin{subequations}
	 	\begin{align}
	 		\eta_0 &= |b|^2 (|c_2|^2 + c_2^* + c_2 + 1), \\
	 		\eta_1 &= (a + 1)^2 |c_2|^2 + (a^2 - 1)(c_2 + c_2^*) + (a - 1)^2, \\
	 		\eta_2 &= (a - 1)^2 |c_2|^2 + (a^2 - 1)(c_2 + c_2^*) + (a + 1)^2, \\
	 		\eta_3 &= (a^2 - 1)|c_2|^2 + (a - 1)^2 c_2^* + (a + 1)^2 c_2 + a^2 - 1, \\
	 		\eta_4 &= 4(|c_2|^2 + 1)\left(h_0 - \frac{J_z}{2}\right) + 8 \operatorname{Re}(c_2(J - iD)).
	 	\end{align}
	 \end{subequations}
	 
	 Figure~\ref{p3}(b) shows the time-dependent ergotropy $\xi$ computed from Eqs.~\eqref{15e} and \eqref{16e} across the three parameter regimes defined in Fig.~\ref{p3}(a). In all cases, ergotropy exhibits oscillatory behavior due to coherent driving, but the maximum value of $\xi$ increases markedly in Regions $R_2$ and $R_3$. This enhancement directly correlates with the growth in energy gap $E_g$, indicating that DMI-induced widening of the spectrum enables greater extractable work. Moreover, the strong charging field ensures that the system can utilize this increased capacity, enabling optimal performance in the DMI-dominated regime.

	\section{Noise resilience in DMI-dominated regimes} \label{sec4}
	
	Having established that DMI enhances the ergotropy of the two-qubit quantum battery, it is pertinent to examine whether such enhancement persists under realistic storage conditions where noise is unavoidable. This question is particularly relevant within the scope of our work, as the ergotropy is shown to vanish asymptotically under repeated application of bit-flip and amplitude damping noise on the single-qubit charged state.
	To explore this, we model the noisy evolution of the charged two-qubit state $\hat{\rho}(t)$ via a quantum channel of the form~\cite{indrajith2021effect}
	\begin{equation}
		\Lambda(\hat{\rho}(t)) = \sum_{i,j} \hat{K}_{ij} \hat{\rho}(t) \hat{K}_{ij}^\dagger,
		\label{17e}
	\end{equation}
	where $\hat{K}_{ij} = \hat{K}_i \otimes \hat{K}_j$ represents the tensor product of single-qubit Kraus operators $\hat{K}_i$ and $\hat{K}_j$, defined in Sec. \ref{sec2}.
	The state $\hat{\rho}(t)$ here denotes the post-charging state of the two-qubit battery, whose stability against decoherence will be assessed under various noise models. In particular, we investigate how the presence of strong DMI influences the system’s robustness against ergotropy degradation, providing insight into the potential operational advantage of such interactions in noisy quantum thermodynamic settings.
	
	\begin{figure}[ht]
		\includegraphics[width=0.8\linewidth]{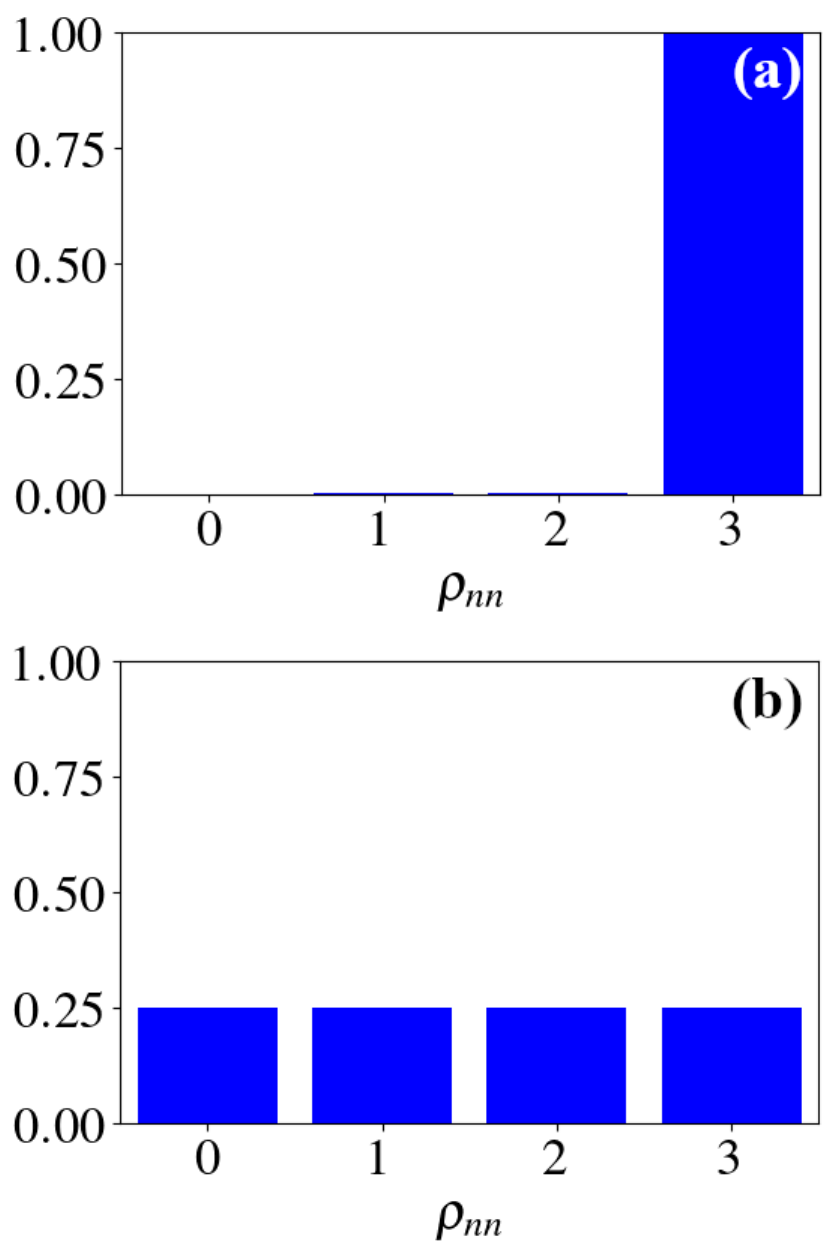}
		\caption{Numerically calculated distribution of the diagonal elements of the density matrix in the asymptotic limit for (a) Amplitude damping noise and (b) Bit-flip noise.}
		\label{p6}
	\end{figure}
	
	\subsection{Amplitude Damping}
	To examine the impact of amplitude damping noise on the stored energy, we analyze the evolution of the density matrix elements under successive applications of the channel defined in Eq.~\eqref{17e}. Let $\rho_{mn}[i]$ denote the matrix elements after $i$ applications of the noise map, i.e., $\Lambda^i(\hat{\rho}(t))$. The evolution of the diagonal elements under amplitude damping is given by
	\begin{subequations}
		\begin{align}
			\rho_{00}[i+1] &= (1-p)^2\, \rho_{00}[i], \\
			\rho_{11}[i+1] &= (1-p)p\, \rho_{00}[i] + (1-p)\, \rho_{11}[i], \\
			\rho_{22}[i+1] &= (1-p)p\, \rho_{00}[i] + (1-p)\, \rho_{22}[i], \\
			\rho_{33}[i+1] &= p^2\, \rho_{00}[i] + p\, \rho_{11}[i] + p\, \rho_{22}[i] + \rho_{33}[i].
		\end{align}
		\label{18e}
	\end{subequations}
	\begin{figure*}[ht]
		\includegraphics[width=1\linewidth]{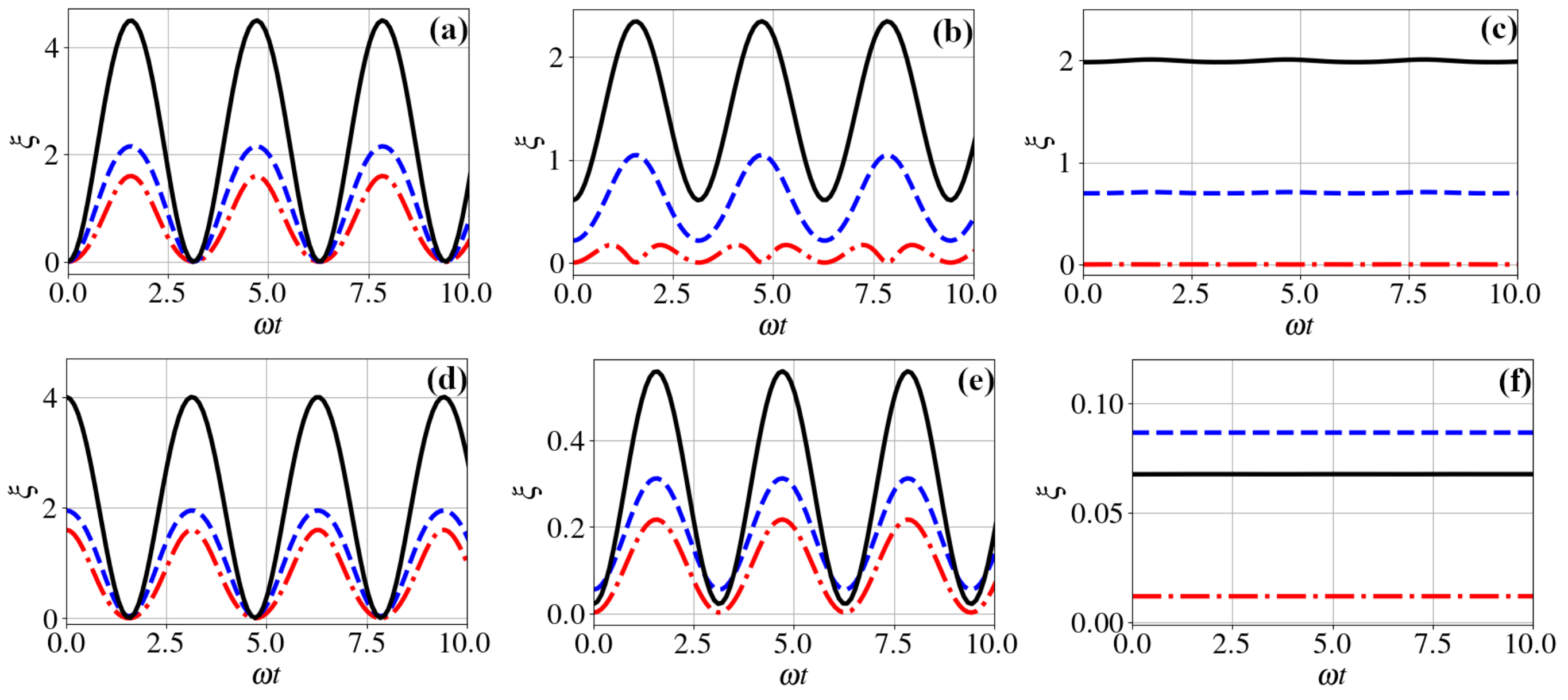}
		\caption{
			Ergotropy $\xi$ of the system after (a) $N=1$, (b) $N=10$, and (c) $N=50$ applications of amplitude damping noise, and (d) $N=1$, (e) $N=10$, and (f) $N=50$ applications of bit-flip noise. 
			The red dash-dotted lines correspond to the $R_1$ regime with $D=0.3$, the blue dashed lines to the $R_2$ regime with $D=1.2$, and the black solid lines to the $R_3$ regime with $D=2.5$. 
			Other parameter values are $J=0.1$, $J_z=0.5$, $\gamma=0.2$, and $p=0.1$.
		}
		
		\label{p4}
	\end{figure*}
	As evidenced from Eq.~\eqref{18e}, the amplitude damping channel progressively transfers the population from the states $\ket{00}, \ket{01}, \ket{10}$ (corresponding to $\rho_{00}, \rho_{11}, \rho_{22}$) into the state $\ket{11}$, represented by $\rho_{33}$. An identical analysis can be performed to show that all the non-diagonal elements of the density matrix decay to zero. With repeated applications, the population accumulates entirely in $\rho_{33}$, leading to the asymptotic state $\Lambda^N(\hat{\rho}(t)) \to \ket{11}\bra{11}$, as illustrated in Fig.~\ref{p6}(a).
	To evaluate the usefulness of this final state from energetic perspective, the asymptotic ergotropy of $\Lambda^N(\hat{\rho})$ can be calculated using \eqref{3e} as
	\begin{equation}
		\lim_{N \to \infty} \xi(\Lambda^N(\hat{\rho}(t))) = 
		\begin{cases}
			\sqrt{D^2 + J^2} + J_z - 1, & \text{if } D > D_c, \\
			0, & \text{otherwise}.
		\end{cases}
		\label{20e}
	\end{equation}
	Here $D_c$ denotes the critical DMI strength separating the passive ($R_1$) and non-passive final ($R_2$ and $R_3$) states.
	
	For $D < D_c$ ($R_1$), the final state $\ket{11}$ is a passive state of the system Hamiltonian $\hat{H}$, resulting in vanishing ergotropy, consistent with the single-qubit case. However, in the regime $R_2$ and $R_3$, the lowest energy state corresponds to the eigenvalue $e_2$ owing to increased asymmetry and hence $\Lambda^N(\hat{\rho})$ is not a passive state. Thus, the residual population in $\ket{11}$ leads to finite ergotropy even in the presence of prolonged noise, highlighting the thermodynamic advantage conferred by the DMI.
	Interestingly, the asymptotic ergotropy is independent of the anisotropy parameter $\gamma$, indicating that in the DMI-dominated regime, asymmetry largely governs the extractable work, rendering symmetric anisotropy  contributions irrelevant in the long-time limit. This result confirms that the ergotropy enhancement achieved through strong DMI persists even during noisy storage, thereby offering robustness in practical settings.

	\subsection{Bit-Flip}
	A similar analysis can be carried out for the bit-flip noise channel. The evolution of the diagonal elements of the density matrix under repeated application of the channel is given by
	\begin{subequations}
		\label{19e}
		\begin{multline}
			\rho_{00}[i+1] = p^2\, \rho_{00}[i] + (1-p)p\, \rho_{11}[i] \\
			+ (1-p)p\, \rho_{22}[i] + (1-p)^2\, \rho_{33}[i],
		\end{multline}
		\begin{multline}
			\rho_{11}[i+1] = (1-p)p\, \rho_{00}[i] + p^2\, \rho_{11}[i] \\
			+ (1-p)^2\, \rho_{22}[i] + (1-p)p\, \rho_{33}[i],
		\end{multline}
		\begin{multline}
			\rho_{22}[i+1] = (1-p)p\, \rho_{00}[i] + (1-p)^2\, \rho_{11}[i] \\
			+ p^2\, \rho_{22}[i] + (1-p)p\, \rho_{33}[i],
		\end{multline}
		\begin{multline}
			\rho_{33}[i+1] = (1-p)^2\, \rho_{00}[i] + (1-p)p\, \rho_{11}[i] \\
			+ (1-p)p\, \rho_{22}[i] + p^2\, \rho_{33}[i].
		\end{multline}
	\end{subequations}

	Unlike amplitude damping, which funnels population into the state $\ket{11}$, the bit-flip channel redistributes population more symmetrically across all diagonal elements. A similar redistribution pattern also appears in the anti-diagonal elements of the density matrix, while all other off-diagonal elements decay under repeated noise. As depicted in Fig.~\ref{p6}(b), the system evolves toward a uniform distribution across the diagonal elements in the long-time limit.
	In this asymptotic regime, the density matrix $\Lambda^N(\hat{\rho})$ takes the form of an X-state \cite{rau2009algebraic},
	\begin{equation}
		\Lambda^N(\hat{\rho}) = 
		\begin{pmatrix}
			0.25 & 0 & 0 & \zeta \\
			0 & 0.25 & \zeta & 0 \\
			0 & \zeta & 0.25 & 0 \\
			\zeta & 0 & 0 & 0.25
		\end{pmatrix},
	\end{equation}
	where $\zeta$ denotes the equally shared anti-diagonal elements. These elements arise from the initial quantum coherence present in the state and are symmetrically distributed due to the bit-flip channel.
	
	The initial coherence originates from the off-diagonal terms proportional to $c_i/(|c_i|^2 + 1)$ in the initial density matrix, where $c_i$ represents either $c_2$ or $c_4$ depending on the parameter regime. The asymptotic value of $\zeta$ is therefore given by
	\begin{equation}
		\zeta = \frac{c_i + c_i^*}{4(|c_i|^2 + 1)} = \frac{\operatorname{Re}(c_i)}{2(|c_i|^2 + 1)}.
	\end{equation}
	The ergotropy of the asymptotic state can then be expressed as
	\begin{equation}
		\lim_{N \to \infty} \xi(\Lambda^N(\hat{\rho}(t))) = \zeta \left[ 2J(\gamma + 1) \mp (L_1 + L_2 - L_3 - L_4) \right],
		\label{21e}
	\end{equation}
	where $L_1$ to $L_4$ are the eigenvalues $e_i$ of the system Hamiltonian $\hat{H}$ arranged in ascending order. The sign in Eq.~\eqref{21e} is chosen such that it matches the dominant eigenvalue of $\Lambda^N(\hat{\rho})$ with the appropriate energy eigenvalue, depending on whether $\zeta$ is positive or negative.
	In the absence of initial coherence (e.g., when $c_2 \approx 0$ or $c_4 \approx 0$), the state evolves into a fully diagonalized maximally mixed state $\mathbb{I}/4$, which is passive with respect to $\hat{H}$ and yields vanishing ergotropy. However, when the initial state contains non-negligible anti-diagonal coherence, the bit-flip noise preserves this structure symmetrically, resulting in finite ergotropy even in the asymptotic limit. The dependence of $\xi$ on $\zeta$ thus highlights the crucial role of initial coherence in enabling noise-resilient energy storage. In particular, the condition $\zeta = 0 \Rightarrow \xi = 0$ clearly illustrates that coherence is necessary for ergotropy preservation under bit-flip noise.

	\subsection{Numerical analysis}

	Due to the analytical complexity of modeling repeated noise in two-qubit systems, we numerically simulate the protocol using the QUTIP library~\cite{johansson2012qutip}. We evaluate the ergotropy of the battery state $\Lambda^N(\hat{\rho})$ using Eq.~\eqref{3e} under successive applications of the amplitude damping and bit-flip channels.
	For the amplitude damping channel, a single application reduces the overall ergotropy slightly compared to the noise-free scenario (Fig.~\ref{p3}(b)), as shown in Fig.~\ref{p4}(a). After 10 applications, the periodicity of ergotropy across parameter regimes persists (Fig.~\ref{p4}(b)), though the $R_1$ regime begins to lose its energy storage capacity. By $N=50$, the ergotropy stabilizes to finite values in the $R_2$ and $R_3$ regimes (Fig.~\ref{p4}(c)), matching the analytical prediction from Eq.~\eqref{20e}. This confirms that in the presence of amplitude damping, the dissipative flow towards the $\ket{11}$ state preserves ergotropy in the DMI-dominated parameter regimes.
	The bit-flip channel, in contrast, not only reduces the ergotropy but also induces a phase shift in the $\xi$ profile even after a single application (Fig.~\ref{p4}(d)). The ergotropy in $R_2$ and $R_3$ remains more resilient compared to $R_1$ for $N=1$ and $N=10$ repeated applications (Figs.~\ref{p4}(d) and (e)). However, $R_2$ and $R_3$ display less robustness under bit-flip noise, indicating a strong dependence on the initial coherence for sustaining nonzero ergotropy. For $N=50$ illustrated in Fig. \ref{p4}(f), the state converges to an X-form with asymptotic ergotropy values well-approximated by Eq.~\eqref{21e}. Unlike amplitude damping, the bit-flip channel does not funnel the population into a single energy level but instead leads to a symmetric redistribution, with the asymptotic ergotropy determined solely by residual coherence in the anti-diagonal elements.
	\begin{figure}[ht]
		\centering
		\includegraphics[width=1\linewidth]{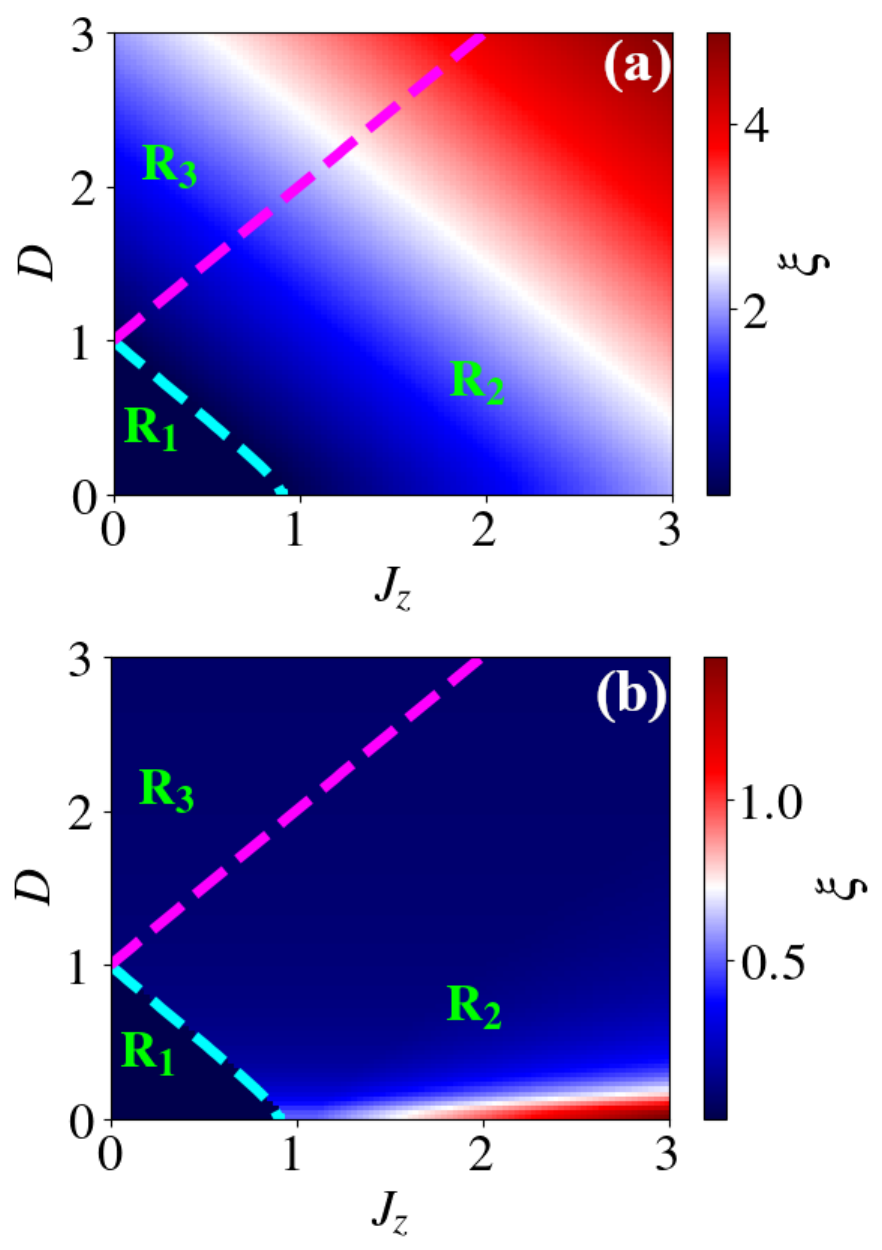}
		\caption{The asymptotic value of ergotropy in parameter space of $(J_z,D)$ for (a) amplitude damping noise and (b) bit-flip noise. Other parameter values are $J=0.1$, and $\gamma=0.2$. Cyan dashed line denotes the boundary between $R_1$ and $R_2$ while the magenta dashed line denotes the boundary between $R_2$ and $R_3$. }
		\label{p5}
	\end{figure}
	
	To better understand the noise resilience across different parameter regimes, Fig.~\ref{p5} shows the asymptotic ergotropy in the $(D, J_z)$ parameter space for both noise types. For amplitude damping [Fig.~\ref{p5}(a)], regime $R_1$ shows complete ergotropy decay, while $R_2$ and $R_3$ maintain robust values. Interestingly, the ergotropy in these regimes is more sensitive to the interaction parameters $D$, $J_z$, and $J$, than to the transition between $R_2$ and $R_3$, indicating that all the interaction parameters govern the asymptotic ergotropy behavior after crossing the $R_1$ regime. Although $R_3$ is the most optimal region for noiseless storage, both $R_2$ and $R_3$ provide resilience under amplitude damping due to the asymptotic steady state being non-passive.
	In the bit-flip case (Fig.~\ref{p5}(b)), regime $R_1$ again proves least effective for long-term storage. However, $R_2$ occasionally shows enhanced performance due to the coherence associated with the $c_2$ term in the initial state. These coherent contributions lead to non-zero anti-diagonal elements in the asymptotic density matrix, sustaining a small yet finite ergotropy. Unlike amplitude damping, the DMI alone is insufficient for maintaining asymptotic ergotropy under bit-flip noise where the coherence in the initial state plays the determining role.
	
	Our results reveal that ergotropy can persist under noise, but the mechanism depends on the type of noise. While the amplitude damping favors population alignment with the asymptotic state under strong asymmetric interactions, the bit-flip noise demands initial coherence. This clarifies that coherence is not merely beneficial for charging \cite{shi2022entanglement} but essential for long-term robustness. In line with earlier studies that emphasized coherence for work extraction \cite{tirone2025quantum}, we show its role in preserving ergotropy during the storage phase of quantum batteries. Overall, our findings improve the understanding of ergotropy dynamics in noisy quantum systems by showcasing how different noise types require distinct ergotropy preservation mechanisms, thereby offering a pathway for noise-resilient ergotropy storage in quantum batteries.
	
	
	\section{Conclusion}\label{sec5}
	
	In this work, we investigated the evolution of ergotropy in quantum batteries subject to repeated noise during the storage phase, focusing on how spin interactions, particularly the Dzyaloshinsky–Moriya interaction can enhance energy robustness. Starting with a single-qubit battery, we established that phase-flip noise selectively preserves ergotropy only for fully charged initial states, while amplitude damping and bit-flip channels monotonically degrade extractable work with repeated exposure.
	
	Extending the analysis to a two-qubit battery governed by the anisotropic Heisenberg XYZ Hamiltonian with DMI, we identified three distinct dynamical regimes determined by critical DMI strengths. In the noiseless case, regions with stronger DMI ($R_2$ and $R_3$) exhibited enhanced ergotropy due to the modified energy spectrum. When noise is introduced, we found that amplitude damping leads to nonzero asymptotic ergotropy in these same regions due to population confinement. The bit-flip noise enables better ergotropy retention in stronger DMI regions for few repeated application but asymptotic ergotropy is preserved only when coherence is embedded in the initial state, resulting in a coherence-dependent reordering of optimal regimes.
	
	These results clarify that both energy-level structure and coherence play complementary roles in mitigating ergotropy decay under noise, depending on the noise channel. In particular, they highlight the potential of chiral spin interactions, such as DMI, as a functional resource for engineering noise-resilient quantum batteries. Our findings suggest that tailoring spin-orbit couplings in few-body systems provides a practical route towards stabilizing extractable work, and motivate future studies into the role of correlations, structured environments, and non-Markovian effects in open quantum thermodynamic settings.
	\onecolumngrid
	\appendix 
	\section{Ergotropy calculation}
	\label{appa}
	
	For a single qubit, ergotropy admits a simplified analytical form derived from Eq.~\eqref{4e}, given by \cite{andolina2019extractable}
	\begin{equation}
		\xi = \frac{1}{2} \left( \rho_{00} - \rho_{11} + \sqrt{(\rho_{00} - \rho_{11})^2 + 4 \rho_{01} \rho_{10}} \right),
		\label{ae1}
	\end{equation}
	where $\rho_{ij}$ denote the elements of the qubit's density matrix $\hat{\rho}$.
	
	\subsection{Phase-Flip Channel}
	
	The action of $N$ successive applications of the phase-flip channel can be captured through the following recursion relations
	\begin{subequations}
		\begin{align}
			\rho_{00}[i] &= \rho_{00}[i-1], \\
			\rho_{01}[i] &= (2p - 1)\, \rho_{01}[i-1], \\
			\rho_{10}[i] &= (2p - 1)\, \rho_{10}[i-1], \\
			\rho_{11}[i] &= \rho_{11}[i-1],
		\end{align}
		\label{ae2}
	\end{subequations}
	with $p$ being the noise parameter.
	By substituting the initial density matrix elements from Eq.~\eqref{2e}, the state after $N$ applications becomes
	\begin{equation}
		\Lambda^{N}(\hat{\rho}) =
		\begin{pmatrix}
			\sin^2{ \omega t} & \frac{-i}{2}(2p - 1)^N \sin{2 \omega t} \\
			\frac{i}{2}(2p - 1)^N \sin{2 \omega t} & \cos^2{ \omega t}
		\end{pmatrix}.
		\label{ae3}
	\end{equation}
	\subsection{Bit-Flip Channel}
	
	The bit-flip noise channel evolves the density matrix via the following recursive equations
	\begin{subequations}
		\begin{align}
			\rho_{00}[i] &= p \rho_{00}[i-1] + (1 - p)\, \rho_{11}[i-1], \\
			\rho_{01}[i] &= p \rho_{01}[i-1] + (1 - p)\, \rho_{10}[i-1], \\
			\rho_{10}[i] &= p \rho_{10}[i-1] + (1 - p)\, \rho_{01}[i-1], \\
			\rho_{11}[i] &= p \rho_{11}[i-1] + (1 - p)\, \rho_{00}[i-1].
		\end{align}
		\label{ae4}
	\end{subequations}
	After $N$ iterations, the matrix elements become
	\begin{subequations}
		\begin{align}
			\rho_{00}[N] &= 
			\left( \sum_{i=0}^{S_1} \binom{N}{2i} p^{N - 2i}(1 - p)^{2i} \right) \mu_2^{2}  + \left( \sum_{i=0}^{S_2} \binom{N}{2i+1} p^{N - 2i - 1}(1 - p)^{2i+1} \right) \mu_1^{2}, \\[1ex]
			\rho_{01}[N] &= 
			\left( \sum_{i=0}^{S_1} \binom{N}{2i} p^{N - 2i}(1 - p)^{2i} \right) \left( -i \,\mu_1\mu_2 \right)  + \left( \sum_{i=0}^{S_2} \binom{N}{2i+1} p^{N - 2i - 1}(1 - p)^{2i+1} \right) \left( i \,\mu_1\mu_2 \right), \\[1ex]
			\rho_{10}[N] &= 
			\left( \sum_{i=0}^{S_1} \binom{N}{2i} p^{N - 2i}(1 - p)^{2i} \right) \left( i \,\mu_1\mu_2 \right) + \left( \sum_{i=0}^{S_2} \binom{N}{2i+1} p^{N - 2i - 1}(1 - p)^{2i+1} \right) \left( -i \,\mu_1\mu_2 \right), \\[1ex]
			\rho_{11}[N] &= 
			\left( \sum_{i=0}^{S_1} \binom{N}{2i} p^{N - 2i}(1 - p)^{2i} \right) \mu_1^{2} + \left( \sum_{i=0}^{S_2} \binom{N}{2i+1} p^{N - 2i - 1}(1 - p)^{2i+1} \right) \mu_2^{2}.
		\end{align}
	\end{subequations}

	Here $\mu_1=\cos{\omega t}$ and $\mu_2=\sin{\omega t}$ with the summation limits given by
	\begin{equation}
		S_1 = 
		\begin{cases}
			\frac{N}{2} & \text{if } N \text{ is even}, \\
			\frac{N - 1}{2} & \text{if } N \text{ is odd},
		\end{cases}
		\qquad
		S_2 = N - S_1.
		\label{ae6}
	\end{equation}
	The binomial coefficient is given by $\binom{N}{i} =   N!/(i!(N - i)!)$.
	
	\subsection{Amplitude Damping Channel}
	
	The amplitude damping channel evolves the density matrix elements according to~\cite{giovannetti2005information}
	
		\begin{align}
			\rho_{00}[i] &= \rho_{00}[i-1] + p \rho_{11}[i-1], \\
			\rho_{01}[i] &= \sqrt{1 - p}\, \rho_{01}[i-1], \\
			\rho_{10}[i] &= \sqrt{1 - p}\, \rho_{10}[i-1], \\
			\rho_{11}[i] &= (1 - p)\, \rho_{11}[i-1].
		\end{align}
	After $N$ applications, the density matrix becomes
	\begin{equation}
		\Lambda^{N}(\hat{\rho}) =
		\begin{pmatrix}
			1 - (1 - p)^N \cos^2{ \omega t}& (1 - p)^{N/2} \left( \frac{-i}{2} \sin{2 \omega t} \right)\\
			(1 - p)^{N/2} \left( \frac{i}{2} \sin{2 \omega t} \right)& (1 - p)^N \cos^2{ \omega t}
		\end{pmatrix}.
		\label{ae8}
	\end{equation}
	By substituting the post-channel density matrix elements into Eq.~\eqref{ae1}, we obtain closed-form expressions for ergotropy after $N$ applications of each noise channel, as analyzed in the main text.
	
	\twocolumngrid
	\bibliography{cite2}

\end{document}